# Spin wave excitations in exchange biased IrMn/CoFe bilayers

Sarah Jenkins , Roy. W. Chantrell, and Richard F. L. Evans

## COLLECTIONS

View Online      Export Citation      CrossMark

---

## ARTICLES YOU MAY BE INTERESTED IN

# Spin wave excitations in exchange biased IrMn/CoFe bilayers

View Online     Export Citation     CrossMark

Sarah Jenkins,[a]  Roy. W. Chantrell, and Richard F. L. Evans[b]

## AFFILIATIONS

Department of Physics, University of York, York YO10 5DD, United Kingdom

**Note:** This paper is part of the special topic on Antiferromagnetic Spintronics.
[a]Author to whom correspondence should be addressed: sarah.jenkins@york.ac.uk
[b]richard.evans@york.ac.uk

## ABSTRACT

Using an atomistic spin model, we have simulated spin wave injection and propagation into antiferromagnetic IrMn from an exchange coupled CoFe layer. The spectral characteristics of the exited spin waves have a complex beating behavior arising from the non-collinear nature of the antiferromagnetic order. We find that the frequency response of the system depends strongly on the strength and frequency of oscillating field excitations. We also find that the strength of excited spin waves strongly decays away from the interfacial layer with a frequency dependent attenuation. Our findings suggest that spin waves generated by coupled ferromagnets are too weak to reverse IrMn in their entirety even with resonant excitation of a coupled ferromagnet. However, efficient spin wave injection into the antiferromagnet is possible due to the non-collinear nature of the IrMn spin ordering.

*Published under license by AIP Publishing.* https://doi.org/10.1063/5.0006232

## INTRODUCTION

Spin wave propagation could potentially be used in the next generation of spintronic devices to transport and process information.[1] These new technologies could massively out perform current devices using electric currents as spin wave propagation occurs at very high frequencies and have a very low energy dissipation. One of the most important issues stopping the development of such devices is tuning the ferromagnetic resonance (FMR) frequency.[2] Ferromagnetic resonance was first predicted by Kittel in 1947.[3] The resonance frequency is experimentally measured by placing the magnetic material in a constant field with strength ($B_0$) and then applying an oscillating field with strength $B_{dr} \exp(i\omega t)$ with an angle of 90° between the fields. When the driving frequency equals the natural precession frequency of the magnetic material, the magnetic material will absorb a large amount of energy from the applied oscillating field, giving a large precession and a much higher measurable magnetization. The resonance frequency $\omega_0$ for a bulk-like sample is given analytically by

$$\omega_0 = \frac{\gamma}{2\pi} \sqrt{B_0(B_0 + B_{ani})}, \qquad (1)$$

where $B_{ani}$ is the anisotropy field of the material and $\gamma = 1.76\,\mathrm{T^{-1}s^{-1}}$ is the gyromagnetic ratio of the electron. In the thin film geometry or elongated nanoparticles, this expression is modified to include the shape anisotropy arising from the long-range dipole–dipole fields.

The FMR frequencies necessary in magnonic devices are beyond 5 GHz, meaning that the FMR has to be increased from the natural resonance state of a material. The FMR frequency can naturally be increased by increasing the uniaxial anisotropy of the material. It has also been shown experimentally that coupling a ferromagnet (FM) to an antiferromagnet (AFM) increases the anisotropy via the exchange bias effect, causing the FMR frequency to increase by up to 10 GHz.[4] Although this shift in resonance frequency has been well known for a number of years, the underlying physical causes and effect on the FM are still poorly understood.[5] Exchange bias occurs when a FM is coupled to an AFM, causing a shift in the magnetic hysteresis loop. The shift is proportional to the interface exchange field of the AFM and caused by a statistical imbalance in the number of spins in each magnetic sublattice.[6] This natural bias field plays the same physical role as anisotropy, while its strength can be manipulated by electrical means.[7,8]

Another aspect of exchange biased systems is the possibility to excite antiferromagnetic spin-wave modes using external magnetic fields. For sufficiently strong coupling, the oscillation of the ferromagnet should excite a dynamic response in the antiferromagnet, possibly inducing higher frequency modes of oscillation. Antiferromagnetic materials have a naturally high resonance frequency, typically in the THz range, and so the FM is not resonating at the natural resonance frequency of the AFM. The nature of the resonant coupling and ability to excite spin wave propagation into an antiferromagnet is poorly studied and of significant practical interest.

In this paper, atomistic simulations of IrMn/CoFe bilayers are used to study the dynamic response of an exchange biased system to an oscillating magnetic field. The exchange bias is found to increase the magnetic resonance frequency of CoFe by about 8 GHz. From this, the change in the magnetization dynamics of CoFe can be observed and the underlying physical mechanisms understood.

## METHODOLOGY

We simulate the system dynamics using an atomistic spin model, where the energy of the system is defined using the spin Hamiltonian,

$$\mathcal{H} = -\sum_{i<j} J_{ij} \mathbf{S}_i \cdot \mathbf{S}_j - \frac{k_N}{2} \sum_{i \neq j}^{z} (\mathbf{S}_i \cdot \mathbf{e}_{ij})^2 - \mu_s \sum_i \mathbf{B}(t) \cdot \mathbf{S}_i, \quad (2)$$

where $\mathbf{S}_i$ is a unit vector describing the spin direction on CoFe/Mn sites $i$, $k_N$ is the Néel pair anisotropy constant, $\mathbf{e}_{ij}$ is a unit vector from site $i$ to site $j$, $z$ is the number of nearest neighbors, and $J_{ij}$ is the exchange interaction. The effective exchange interactions ($J_{ij}$) were limited to nearest and next nearest neighbors for the antiferromagnet[9,10] and nearest neighbors for the ferromagnet. The interfacial exchange is set to a quarter of the bulk exchange in the antiferromagnet guided by first principles calculations.[11,12] These parameters are outlined in Table I. We note that we omitted the dipole contribution to the Hamiltonian in our simulations for reasons of computational efficiency. In addition, given the small size and cubic shape of our system, the dipole field contribution to the spin wave excitations would be negligible.

**TABLE I.** Parameters used in the atomistic spin model. NN is short for the nearest neighbor and NNN is short of the next nearest neighbor.

| Quantity | Material | Symbol | Value | Units |
|---|---|---|---|---|
| NN exchange | IrMn | $J_{ij}^{nn}$ | $-6.4 \times 10^{-21}$ | J |
| NNN exchange | IrMn | $J_{ij}^{nnn}$ | $5.1 \times 10^{-21}$ | J |
| Interlayer exchange | ... | $J_{ij}$ | $1.5 \times 10^{-21}$ | J |
| NN exchange | CoFe | $J_{ij}$ | $4.6 \times 10^{-21}$ | J |
| Neel pair anisotropy | IrMn | $k_N$ | $-4.2 \times 10^{-22}$ | J |
| Anisotropy constant | CoFe | $k_u$ | 0 | J |
| Magnetic moment | IrMn | $\mu_S$ | 2.6 | $\mu_B$ |
| Magnetic moment | CoFe | $\mu_S$ | 2.5 | $\mu_B$ |
| Gilbert damping | Both | $\lambda$ | 0.05 | ... |

We simulate the dynamic behavior using the stochastic Landau–Lifshitz–Gilbert (sLLG) equation applied at the atomistic level[13,14] and given by

$$\frac{\partial \mathbf{S}_i}{\partial t} = -\frac{\gamma}{1+\lambda^2} [\mathbf{S}_i \times \mathbf{B}_{eff} + \lambda \mathbf{S}_i \times (\mathbf{S}_i \times \mathbf{B}_{eff})], \quad (3)$$

where $\lambda$ is the Gilbert damping constant and $|\gamma|$ is the gyromagnetic ratio. The effective field $\mathbf{B}_{eff}$ is calculated from the derivative of the spin Hamiltonian with respect to the local spin moment ($\mathbf{B}_{eff} = -\mu_S^{-1} \partial \mathcal{H} / \partial \mathbf{S}_i + \mathbf{B}_{th}^i$), where $\mathbf{B}_{th}^i = \Gamma(t) \sqrt{\frac{2\lambda k_B T}{\gamma \mu_s \Delta t}}$ and $\Gamma$ is a 3D random Gaussian distribution. The sLLG equation is integrated using a second order predictor corrector Heun scheme.[13] All simulations were conducted at $T = 0$ K to eliminate thermal noise. The calculations have been carried out using the VAMPIRE software package.[13,15]

The system is oriented with the (111) crystal direction perpendicular to the plane as is common in exchange biased devices, shown schematically in Fig. 1(a). Each of the different colored spheres represents a different atomic site of the four total in the FCC base crystal unit cell. For the disordered $\gamma$-IrMn$_3$ phase, the Ir atoms are randomly allocated to each of the sites (not shown). Periodic boundary conditions are applied along the $x$, $y$ directions of the system to avoid missing exchange and Néel pair anisotropy bonds at the edges. To set the exchange bias, we use an algorithmic setting procedure to determine the orientations of each magnetic sublattice for a given bias direction and pick the configuration with the lowest energy. To verify that the exchange bias is set correctly, we perform a slow hysteresis loop with critical damping ($\lambda = 1$) and find a shift of $-0.15$ T at $T = 0$ K with unconditional stability at zero external applied field, as shown in Fig. 1(b). The exchange bias in our model comes from a statistical imbalance in the number of spins in each magnetic sublattice, leading to a small and realistic exchange bias field in comparison with experiments.[17,18] Our model also correctly reproduces the tetrahedral (3Q) and triangular (T1) ground states for the disordered and ordered IrMn$_3$ phases, respectively,[9] shown schematically in Fig. 1(b).

To explore the dynamics of a coupled ferromagnet and antiferromagnet, we model a single grain of $\gamma$IrMn$_3$ coupled to an effective ferromagnet with magnetic properties similar to Co$_{40}$Fe$_{60}$ with zero magnetocrystalline anisotropy similar to the bulk CoFe alloy.[19] At this composition, the CoFe alloy has an isotropic point which is useful for magnetic sensors using an exchange biased system to avoid an anisotropic bias in the magnetic orientation of the film. To determine the response of the IrMn/CoFe system to spin wave excitations, we simulated a ferromagnetic-resonance type experiment by applying an oscillating magnetic field $\mathbf{B}_f$ along the $y$-direction perpendicular to the exchange bias direction[20] set along $x$. Note that, here, we mostly consider a zero static-field resonance where the external field is set to zero. The time-dependent applied magnetic field is given by

$$\mathbf{B}(t) = \mathbf{B}_0 + \mathbf{B}_{fmr} \sin(2\pi\omega t), \quad (4)$$

where $\omega$ is the driving frequency varied in the range $\omega = 0.1 - 100$ GHz.

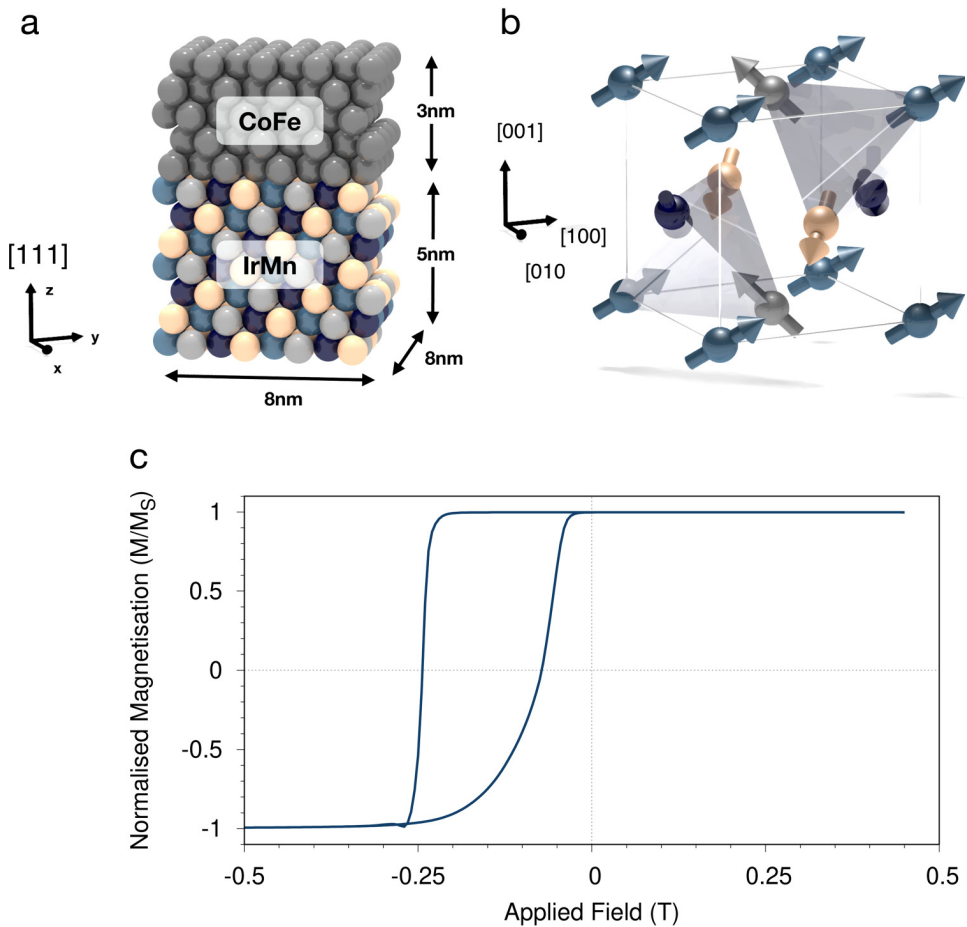

**FIG. 1.** (a) Schematic diagram of (111) oriented CoFe/γ-IrMn₃ bilayer. (b) Simulated ground state magnetic structure of γ-IrMn₃. (c) Simulated hysteresis loop showing exchange bias field arising from an imbalance of sublattice spins at the interface.[16]

In our simulations, we consider only a small sample of a large continuous film and also include periodic boundary conditions. While we naturally include the exchange bias effect at the single grain level, the small system size means that spin waves are naturally limited to an 8 nm wavelength, therefore excluding long-wavelength excitations. In general, such excitations are important and are driven by inhomogeneities in the magnetic properties in the film, leading to incoherent spin wave excitations. In a polygranular exchange biased system, we would expect that the distribution of exchange bias in the film will lead to the appearance of lower energy long-wavelength modes, which are neglected in our present simulations. Such simulations are computationally expensive due to the atomistic nature of our model, and so here we focus on the intrinsic coherent mode excitations in an exchange biased system.

## FERROMAGNETIC RESONANCE IN EXCHANGE BIASED SYSTEMS

Initially, we consider the frequency response of CoFe without coupling it to an AFM. We approximate the exchange bias field as a 0.1 T static field along $y$, $\mathbf{B}_0 = \mu_0 H_{ex}$. We chose this value because it matches the exchange bias calculate in our IrMn/CoFe bilayer. We consider the response of only a single FM spin of CoFe, where the exchange interactions and dipole fields are not included. An oscillating field was applied along the $x$ direction with strength 0.01 T and a driving frequency of 10 GHz. We can predict the resonance frequency for our FM using Eq. (1) giving $\omega_0 = 2.8$ GHz. The response of the ferromagnet is shown in Fig. 2(a). The response shows a sinusoidal oscillation with a single frequency. Unlike in the non-zero field FMR experiment, the resonance monotonically decreases with driving frequency, since low frequencies give sufficient time for the ferromagnetic magnetization to completely align with the field direction. At a 10 GHz driving frequency the ferromagnetic layer shows a relatively weak response perpendicular to the driving field.

The simulated resonance peak is shown in Fig. 2(b), with a large peak at 2.8 GHz as expected. The spectrum can be exactly fitted by a Lorentzian curve

$$L(\omega) = \frac{A}{\pi} \frac{0.5f}{(\omega - \omega_0)^2 + (0.5f)^2} \quad (5)$$

where $f$ represents the width of the curve and $A$ its amplitude.[20]

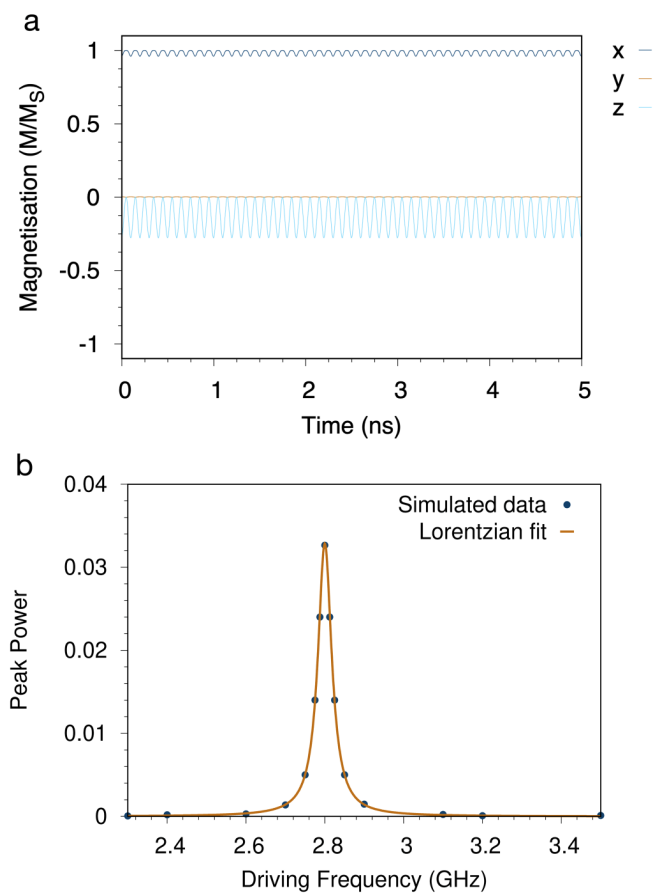

**FIG. 2.** Ferromagnetic resonance for an isolated layer of CoFe (a) showing a sinusoidal variation of the magnetization components in time. (b) The frequency dependent resonance curve gives a peak with a driving frequency of 2.8 GHz as predicted from Eq. (1), and a fit to a Lorentzian curve given by Eq. (5).

Our results for an isolated CoFe layer agree with known analytical results demonstrating the basic correctness of our numerical model. This result would also be representative of a simple micromagnetic calculation, with no simulated degrees of freedom from the antiferromagnet and a simple fixed magnetic field.

We next consider a more realistic situation coupling the CoFe layer to an antiferromagnet with a resulting exchange bias field. Here, the simulation is slightly modified, removing the 0.1 T external field (representing the exchange bias) but the oscillating field remained the same with a fixed driving frequency of $\omega = 10$ GHz. The time-dependent response of the $x$-component of the magnetization of the ferromagnetic layer as a function of different oscillating field strengths is shown in Fig. 3 (left) and the associated Fourier transform (right). The coupling of the ferromagnetic layer to the $\gamma$-IrMn$_3$ film leads to a much richer time-dependent behavior in the case of strong excitations and large excursions of the magnetization. The Fourier transforms clearly show the evolution of a single main frequency component in Fig. 3(a) to a larger

10 GHz frequency response with a superimposed high frequency component in Fig. 3(b) to a complex multi-frequency response in Fig. 3(c). The complex response of the system to different excitation field strengths suggests that the response of a exchange biased system is not at all trivial. Indeed, for the strongest excitation field of $B_{fmr} = 0.1$ T shown in Fig. 3(c), a beating-like behavior of the magnetization is observed.

At this point, it is useful to consider the atomic origin of the exchange bias effect in CoFe/IrMn systems, caused by a natural imbalance in the number of interface spins in one of the four magnetic sublattices. When these interface spins are coupled to a neighboring ferromagnet by the exchange interaction, the ferromagnet develops a strong preference for aligning with the net interface magnetization, which is also strongly exchange coupled to the bulk of the antiferromagnet. While the exchange bias effect on the ferromagnet is well understood, the reciprocal effect on the antiferromagnet is often neglected. However, the static evolution of the reversible and irreversible components of the interfacial spins is known to follow the orientation of the ferromagnet.[17] Given the different atomic environment of the interfacial spins, it is reasonable to assume that the natural resonance frequency of those spins is quite different from the bulk of the ferromagnetic layer and also from the bulk antiferromagnet. We can then consider our exchange bias system as a system of coupled oscillators with different natural frequencies. This naturally explains the appearance of complex beating and also its semi-chaotic behavior in time. The natural frequency of optical excitations of antiferromagnets is in the THz range, and so here we only excite coherent (acoustic) modes in the antiferromagnet, where the 3Q ground state is rotated coherently in space, equivalent to rotation of a ferromagnet in an anisotropy field. Of course, the differing properties of the interface and bulk of the antiferromagnetic layer mean that the response of the antiferromagnet is not entirely coherent, contributing to the broad frequency response in the strongly excited case. There may also be contributions to the beating behavior due to the finite size of the simulated grain, leading to standing spin waves in the system. However, this would be peculiar to non-collinear antiferromagnets as no such effects are seen in ferromagnets where strong exchange interactions give a uniform response. Further optical FMR measurements may yield observations of more complex frequency responses in the case of high frequency excitations.[21] We note that chaotic-like dynamic behavior has also been observed before in purely ferromagnetic systems[32] under certain excitation conditions. Under the excitation conditions simulated here, this never appears for the purely ferromagnetic case, and so the complex dynamics are a clear result of the coupling to the antiferromagnet.

We now consider the frequency dependence of the response of the ferromagnet for a fixed driving field strength of $|\mathbf{B}_{fmr}| = 0.05$ T, shown in Fig. 4. As with the field strength dependent response, there is a complex frequency dependence of the power. For a driving frequency $\omega = 10$ GHz close to resonance, the frequency response is extremely broad, with moderate excitations at all frequencies, and a broad peak around 10 GHz. In contrast, an off-resonant excitation at $\omega = 1$ GHz shows a clearer spectral character, with a principle peak at the driving frequency, and an exponential decrease in the power at harmonic frequencies at 2, 3, 4, and 5 GHz. The presence of harmonics gives a square-wave character to the time series, but the

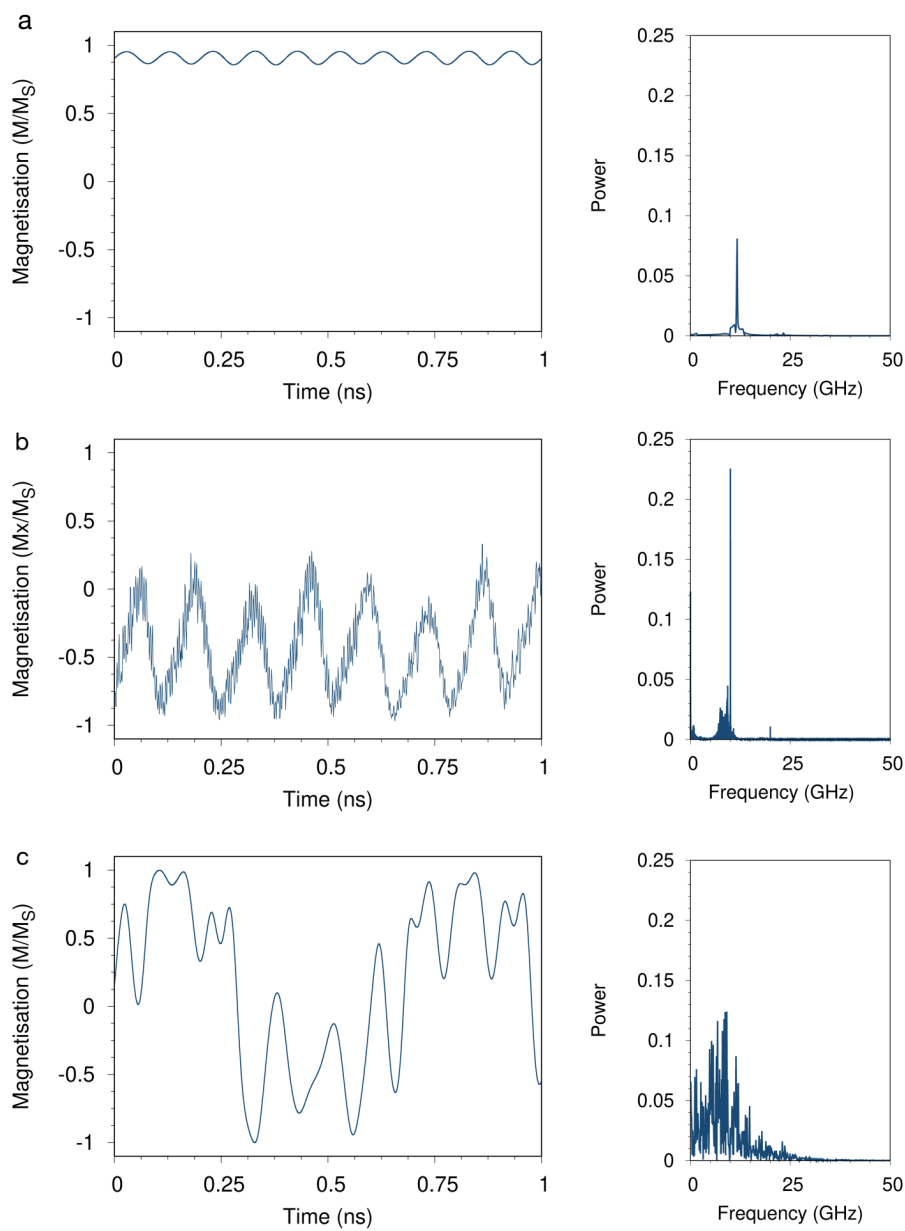

**FIG. 3.** Response of the ferromagnetic layer to a 10 GHz driving field frequency with an applied field strength of (a) $B_{fmr} = 0.01$ T, (b) $B_{fmr} = 0.05$ T, and (c) $B_{fmr} = 0.1$ T. The time data are offset by $t = 50$ ns after the simulation is started to remove transient effects.

logarithmic decay of the power of the higher harmonics makes this a weak effect giving a small asymmetry to the single frequency sine wave. There is an additional weak frequency component in the 10–14 GHz range close to the natural resonance, which is not visually apparent but shows weak excitation of the antiferromagnet at the 1 GHz driving frequency. For higher frequency driving fields at 5–50 GHz, the harmonic components persist at integer multiples of the driving frequency, with the same logarithmic drop-off of the power with increasing harmonics, characteristic of a near-undetectable asymmetry in the $m_x(t)$ response. Overall, our data suggest that the complex frequency response and beating effects only occur when the driving frequency is only close to the resonance frequency of around 10 GHz.

The frequency dependence of the peak power (principal peak) as a function of the driving frequency is shown in Fig. 5 for both the FM and AFM at the interface, showing the strength of the principal excitation as well as the transmission of power into the interfacial layer of the AFM. The power in the FM shows a broad peak in the vicinity of the resonance frequency around 10 GHz but with a much larger linewidth than that seen for the isolated layer in Fig. 2. The increase in linewidth is indicative of an increase in damping, as previously seen in exchange biased systems.[23]

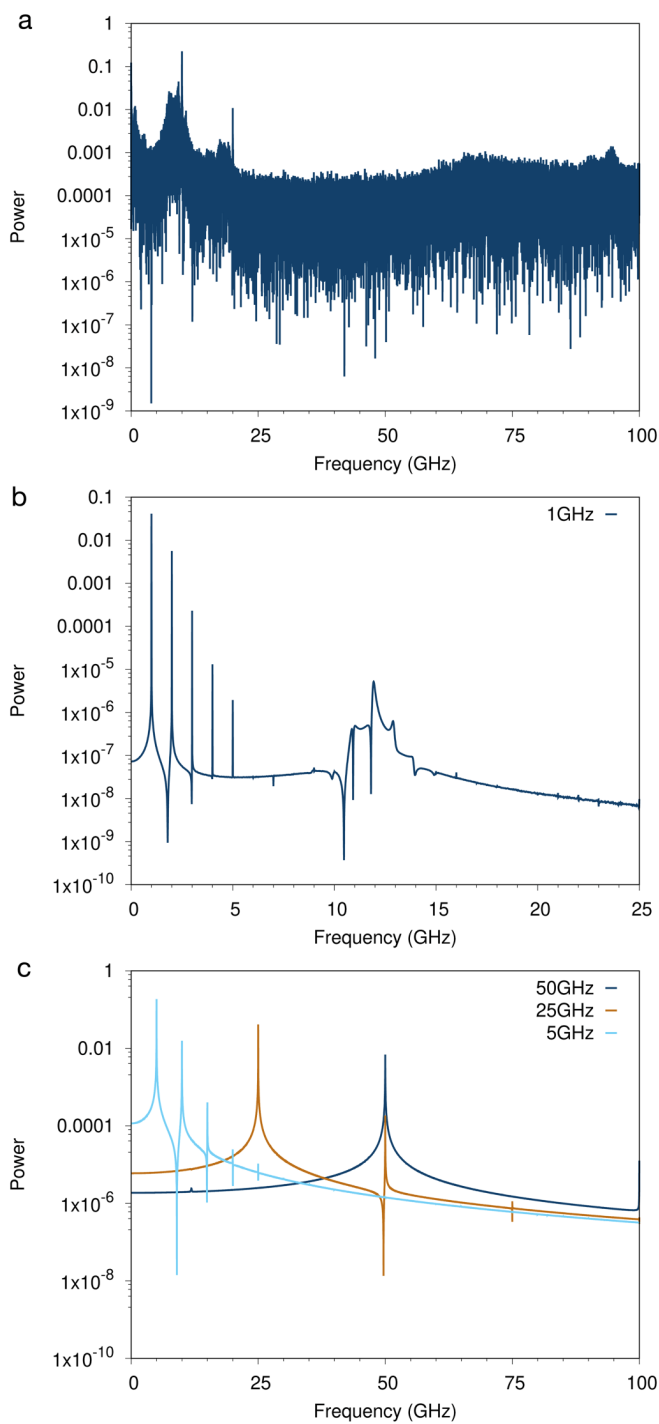

**FIG. 4.** Response of the ferromagnetic layer to a 0.05 T oscillating field with a driving frequencies of (a) 10 GHz, (b) 1 GHz, and (c) 5, 25, and 50 GHz. The power is calculated from a Fourier transform of the time-dependent magnetization after 50 ns, plotted on a logarithmic scale to accentuate the different excited frequencies.

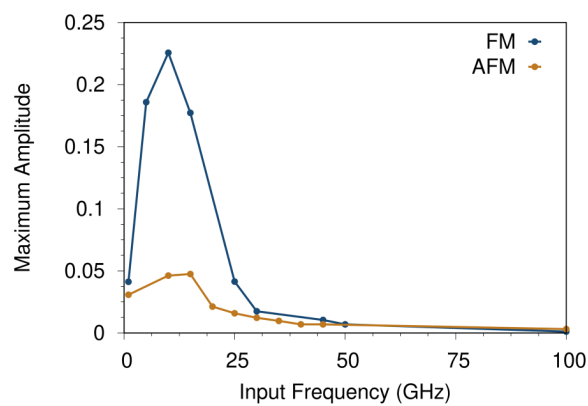

**FIG. 5.** Comparative plot of the peak power spectral density for the FM and for the AFM interface (0 nm from the interface) as a function of different input driving frequencies for CoFe/γ-IrMn₃ with the 0.05 T driving field.

The excitation peak power in the interfacial AFM layer follows a similar trend to the FM, though with a much broader peak, indicating the propagation of spin waves into the AFM across the sub-20 GHz range of excitation frequencies.

In exchange biased systems, the exchange bias field is dependent on the thickness of the FM layer, as the effect is purely interfacial origin, at least for thin films where exchange coupling dominates and the FM behaves coherently. In Fig. 6, we present the calculated power spectrum for a 0.05 T oscillating field at a driving frequency of 10 GHz for different ferromagnet thicknesses. As expected, there is a strong peak at the driving frequency of $\omega = 10$ GHz. However, the complex character of the frequency response of the FM for the 2 nm thick film is strongly suppressed. As the exchange bias field is reduced with increasing FM thickness, this leads to weaker coupled excitations. The principal peak at 10 GHz systematically decreases in height and increases in breadth with increasing film thickness and becomes spectrally cleaner at intermediate off-resonant frequencies, showing an absence of significant beating. In addition, a weak secondary peak appears at 3.3 GHz and 4.5 GHz for 3 nm and 4 nm thicknesses, respectively, that is clearly thickness dependent, which could indicate a shift of the natural resonance of the system. Overall, this suggests that with increasing thickness the coupling between the FM and AFM reduces the strength of excitations and with it a less complex coupled excitation.

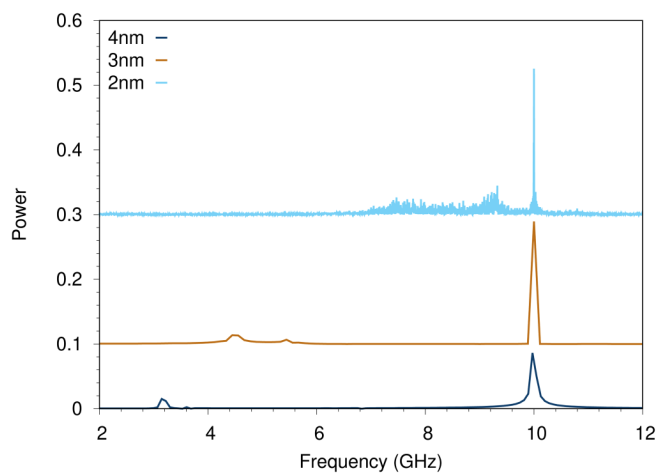

**FIG. 6.** Thickness dependence of the excited spin wave spectrum for a 0.05 T oscillating field at a driving frequency of 10 GHz. With increasing film thickness, the spin wave spectrum develops a pure spectral character to the decreased strength of the coupling.

## EXCITATION OF SPIN WAVES IN THE ANTI-FERROMAGNET

The coupling of the ferromagnet to the antiferromagnet naturally leads to the propagation of spin waves into the IrMn layer, known as evanescent AFM spin wave modes.[24] The interfacial nature of exchange bias coupling naturally leads to a stronger excitation at the interfacial layer, while the strong magnetic anisotropy in IrMn reduces the strength of excitations away from the interface. The time-dependent response of the dominant magnetic sublattice at the interface (responsible for the exchange bias) and bottom of the IrMn layer is shown in Fig. 7. The data show a similar characteristic beating behavior but much weaker than observed for the ferromagnet, even at the interfacial layer in direct contact with it. This suggests that the coupling between the antiferromagnet and ferromagnet is far from rigid, despite the strong exchange bias field and unidirectional coupling, and it seems that high frequency excitations allow for large excursions of the ferromagnet from the biasing antiferromagnet. Naturally, the IrMn moments are highly stable due to their high magnetic anisotropy.

To further explore the spectral characteristics of the response of the antiferromagnet, we computed a Fourier transform of the equilibrated signal taken over a period of 100 ns and shown in Fig. 8 for the dominant IrMn magnetic sublattice at (0 nm) and far away (3.6 nm) from the interface. The spectral characteristics of the antiferromagnetic sublattice at the different locations are very similar, exhibiting a principal peak at the 10 GHz driving frequency, but also with significant peaks with significant bandwidth. In particular, there are strong oscillations in the 100 GHz range characteristic of the naturally fast dynamics of antiferromagnetic materials. In the vicinity of the driving frequency at 10 GHz, the spectral power density is approximately half as strong for the bottom IrMn layer compared to the interface, while in the 100 GHz region, this is reduced to approximately 1/3. This suggests a natural frequency dependence of the attenuation of spin waves propagating in the antiferromagnet, with higher frequency modes being dissipated more strongly. Interestingly, the data show a significant coupling of spin waves between the ferromagnet and antiferromagnet, and a similar behavior has been observed in recent experiments for the FeMn/Py system[25] with broad implications for tuning the dynamic response of materials in the 5–200 GHz frequency range.

Although the excited spin waves in the antiferromagnet have an intrinsic frequency dependent transmission coefficient, the coupled system itself also has a natural resonance at around 10 GHz. Thus, at the couple resonance frequency, the interfacial spins are strongly excited, leading to larger excitations in the antiferromagnet. To investigate the propagation of spin waves into the antiferromagnet, we have computed the strength of excitation as a function of different driving frequencies. The decay of the spin wave magnitude as a function of distance from the interface and driving frequency is shown in Fig. 9. The magnitude of the excitations shows an exponential-like decay away from the interface but with a slow tail suggesting that spin waves excited in the antiferromagnet will propagate a significant distance before decaying to an undetectable level, perhaps up to tens of nanometers. To characterize the frequency dependence of the decay of spin waves in the antiferromagnet, we fit a general two-term exponential function of the form

$$f(x) = A_1 \exp(-b_1 x) + A_2 \exp(-b_2 x), \qquad (6)$$

where $A_1$, $A_2$, $b_1$, and $b_2$ are fitting parameters. The fitted characteristic decay lengths are shown in Table II. The decays are poorly fitted with a single exponential, indicating two important length scales associated with the decay of the spin waves. The strongest excitation is clearly in the 10 GHz frequency region, while 1 GHz and 100 GHz are both weaker being off-resonant. From the fitted values, it is clear the near interface decay is extremely rapid in all three cases, with a characteristic length scale of around 3 Å or approximately one atom. In contrast, the long-distance decay is more strongly dependent on the driving frequency, showing a systematic reduction of the propagation distance with increasing driving frequency. These two characteristic lengthscales can then be associated with two different physical phenomena. The short-range decay is due to the interfacial atomic layer where there is a large reduction of magnetic anisotropy in IrMn due to the missing exchange and pair anisotropy bonds. This allows for relatively large excursions of the coupled interfacial spins under ferromagnetic resonance, as well as spins in the IrMn sublayer that are directly exchange coupled to the interface spins. The second characteristic decay is then due to the propagation of excitations into the bulk of the antiferromagnet, where the exchange coupling and magnetic anisotropy are both large. Here, the excitations are coherent in nature, exciting collective (acoustic) oscillations of the antiferromagnetic ground state structure. Here, lower frequency excitations are preferred since the response of the antiferromagnet becomes quasi-static, while at higher driving frequencies, the antiferromagnet is less able to respond to the higher frequency oscillations. This explains the much longer decay length for the off-resonant excitation at 1 GHz, even though the response of the ferromagnet is weaker. This supports the idea that the frequency overlap of the ferromagnet and antiferromagnet response is essential in correctly propagating spin waves into a coupled antiferromagnet.[25] The resonance in the 10 GHz range is close to reported experimental values for a similar material system.[26]

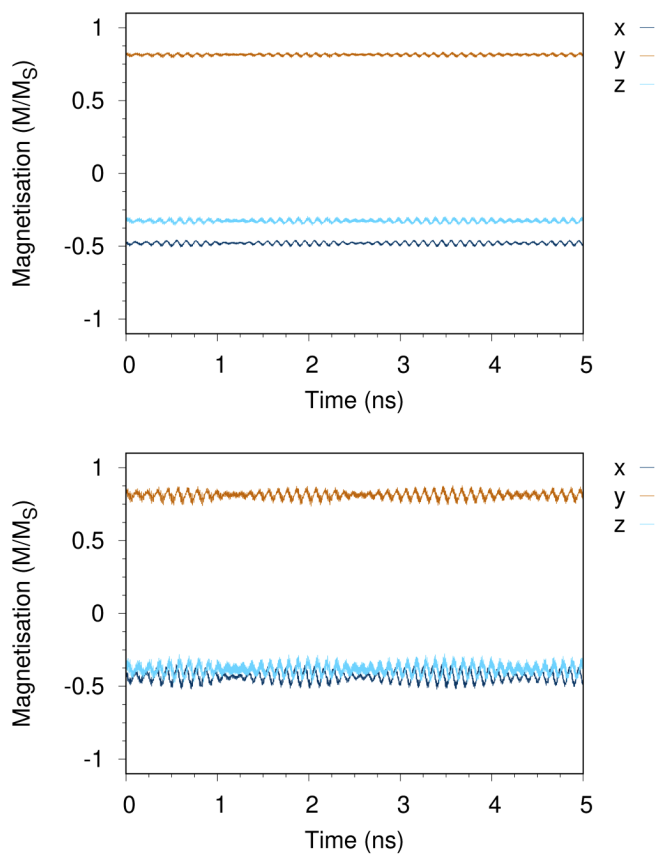

**FIG. 7.** Time-dependent sublattice magnetizations of the dominant antiferromagnetic sublattice (responsible for the exchange bias) at the interface (top) and bottom of the IrMn layer (bottom) for a driving frequency of 10 GHz. The time data are offset by $t = 50$ ns after the simulation is started to remove transient effects.

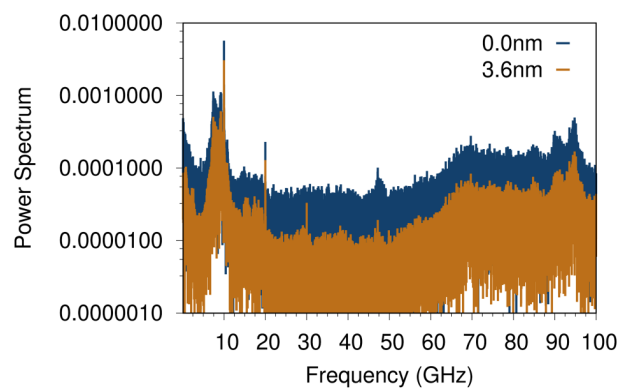

**FIG. 8.** The frequency spectra for the AFM at the interface (0.0 nm from the interface) and at the edge (3.6 nm from the interface) obtained from the Fourier transform of the FMR data.

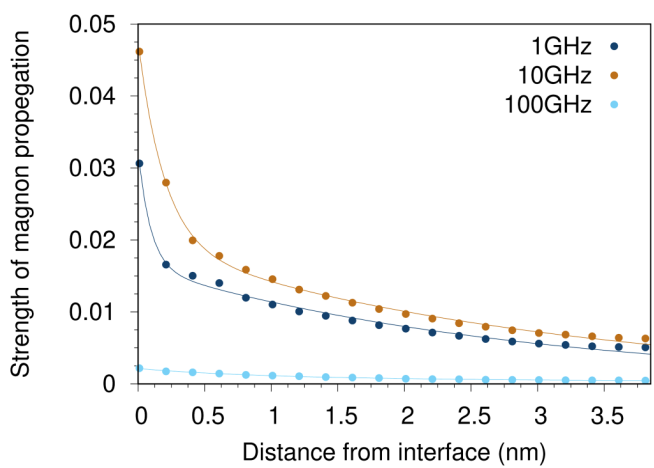

**FIG. 9.** Layerwise response of IrMn. The maximum response of a sublattice for each layer of the AFM. The highest amplitude frequency response occurs at zero distance from the interface. As the distance from the interface increases the amplitude decreases. This happens for all input frequencies.

**TABLE II.** Fitted frequency decay lengths in the antiferromagnet from Eq. (6) for different excitation frequencies.

| Frequency (GHz) | Decay length (nm) | |
| --- | --- | --- |
| | $b_1$ | $b_2$ |
| 1 | $0.356 \pm 0.016$ | $10.673 \pm 1.812$ |
| 10 | $0.333 \pm 0.013$ | $5.115 \pm 0.263$ |
| 100 | $0.179 \pm 0.094$ | $1.089 \pm 0.237$ |

## CONCLUSIONS

In conclusion, we have performed atomistic modeling of spin wave injection and propagation into IrMn from an exchange biased CoFe-bilayer. We have found that the spectral characteristics of the exited spin waves have a complex behavior and systematic beating arising from the non-collinear nature of the antiferromagnetic order, behaving as coupled oscillators when strongly excited. The strength of excited evanescent spin waves decays strongly away from the interfacial layer but with a slow tail, and high frequency waves in the 100 GHz range are more strongly attenuated. Our results suggest that efficient spin wave injection into IrMn is possible and that the combined resonance frequency is significantly higher than for the isolated ferromagnetic layer. Furthermore, the non-collinear nature of the antiferromagnetic order enables the efficient excitation of antiferromagnetic spin waves in the 1–20 GHz frequency range, which may enable new applications and devices with a tunable frequency response. Here, we neglect the role of long-range dipole–dipole interactions on computed spin wave spectra, instead focusing on the fundamental small-scale interactions in an exchange biased system. For extended systems, the dipole–dipole interactions are expected to make an important contribution and will be necessary for a detailed comparison with experimental data. Future work will consider the effects of temperature, long-distance decay of the generated spin waves, and additional composition and ordering effects.

## AUTHORS' CONTRIBUTIONS

S.J. performed the atomistic simulations, analyzed, and plotted the data. S.J. and R.F.L.E. drafted the paper. All authors contributed to the writing of the manuscript and interpretation of results.

## ACKNOWLEDGMENTS

We gratefully acknowledge the provision of computer time made available on the VIKING cluster, a high performance compute facility provided by the University of York. The authors are grateful to Vincent Baltz for insightful discussions.

## DATA AVAILABILITY

The data that support the findings of this study are available from the corresponding author upon reasonable request.